\documentclass[11pt]{article}
\usepackage{amstex,psfig}
\newcommand{\ol}{\setlength{\itemsep}{0pt.}\begin{enumerate}}
\newcommand{\eol}{\end{enumerate}\setlength{\itemsep}{-\parsep}}
\newcommand{\remove}[1]{}
\newcommand\nd{\noindent}
\newtheorem{thm}{Theorem}
\newtheorem{prop}[thm]{Proposition}
\newtheorem{lemma}[thm]{Lemma}

\newcommand{\QED}{\hfill$\Box$}
\newcommand{\qed}{\hfill$\Box$}
\newcommand{\proof}{\noindent {\bf Proof}\ \ }
\newcommand{\ber}{{\begin{eqnarray*}}}
\newcommand{\eer}{{\end{eqnarray*}}}
\newcommand\bc{{\bf c}}
\newcommand{\bb}{{\bf b}}
\newcommand{\be}{{\bf e}}

\newcommand{\bs}{{\bf s}}
\newcommand{\bu}{{\bf u}}
\newcommand{\bv}{{\bf v}}
\newcommand{\bw}{{\bf w}}
\newcommand{\bx}{{\bf x}}
\newcommand{\by}{{\bf y}}
\newcommand{\bz}{{\bf z}}
\def\wt{\qopname\relax{no}{wt}}
\def\supp{\qopname\relax{no}{supp}}
\def\tr{\qopname\relax{no}{Tr}}
\def\dim{\qopname\relax{no}{dim}}
\def\dist{\qopname\relax{no}{dist}}
\def\Pr{\qopname\relax{no}{Pr}}
\def\im{\qopname\relax{no}{im}}
\newcommand{\ten}{\otimes}
\newcommand\Proof{\noindent{\sc Proof. }}
\newcommand{\pll}{\propto}

\begin{document}

\title       {Quantum Error Detection I: Statement of the Problem}

\author {Alexei Ashikhmin
\thanks {Los Alamos National Laboratory, Group CIC-3, Mail Stop P990, 
Los Alamos, NM 87545. 
%{E-mail: alexei@c3serve.c3.lanl.gov}
} \and
Alexander  Barg
\thanks{Bell Laboratories, Lucent Technologies, 
  600 Mountain Avenue 2C-375,
  Murray Hill, NJ 07974.
%{E-mail: abarg@research.bell-labs.com}
} \and
Emanuel  Knill
\thanks { Los Alamos National Laboratory 
  Group CIC-3, Mail Stop P990, 
Los Alamos, NM 87545. 
%{E-mail: knill@c3serve.c3.lanl.gov}
} \and 
Simon Litsyn
\thanks{ Department of Electrical Engineering-Systems, Tel Aviv University, 
Tel Aviv 69978, Israel. 
%{E-mail: litsyn@eng.tau.ac.il}
}
}

\date{}
\maketitle

\begin{abstract}
This paper is devoted to the problem of error detection with
quantum codes. In the first part we examine possible problem settings
for quantum error detection. Our goal is to derive a functional
that describes the probability of undetected error under 
natural physical assumptions concerning transmission 
with error detection with quantum codes. We discuss
possible transmission protocols with stabilizer
and unrestricted quantum codes. The set of results proved in 
part I shows that in all the cases considered the average probability
of undetected error for a given code is essentially given by one 
and the same function of its weight enumerators. This enables us
to give a consistent definition of the undetected error event.
In part II we derive bounds on the probability of undetected error 
for quantum codes.

In the final section of the paper we examine polynomial invariants
of quantum codes and show that Rains's ``unitary weight enumerators''
\cite{ref 2} are known for classical codes under the name
of binomial moments of the distance distribution.
As in the classical situation, they provide an alternative expression
for the probability of undetected error.
\end{abstract}


\section{Introduction}
The possibility of correcting decoherence errors in entangled 
states was discovered by Shor in \cite{shor1} and
Steane \cite{steane1}. Since then the theory of quantum codes has 
been a topic of intense study. Error processes in the
depolarizing channel are characterized in \cite{knill} where it was 
shown that one can restrict attention to error operators
given by Kronecker products of Pauli matrices. This opened
the possibility of finding parallels between the theory
of classical error correcting codes and their quantum
counterparts.

This theory has developed simultaneously in different, often
not very related directions. 
Mathematically one of the most interesting 
works on quantum codes is the paper by 
Calderbank et al. \cite{Cal gf}, where the authors introduce
a class of so-called stabilizer quantum
codes (independently found in \cite{Gottesman}), 
and discovered a beautiful geometric and group-theoretic
connection that shows that such codes can be studied using 
the well-developed theory of classical codes over $GF(4).$ 
Stabilizer codes share some of the properties of classical
linear codes.

One of the most important concepts in classical coding
theory is the notion of decoding, decoding error, and
the probability of this event for a given (random) error
process in the channel. In the quantum case a similar theory
still awaits construction; even defining decoding does
not seem an easy problem. In the classical coding theory
the simplest known decoding algorithm is testing
the received vector for containment in the code; if the
test fails, the decoder detects an error. Clearly,
the error will not be detected if the sum of the transmitted 
vector and the error is itself a code vector.

The focus of our paper is error detection
by quantum stabilizer and nonstabilizer codes. 
To give an analogy, let us recall the definition of error 
detection for classical codes.
Let $D$ be a code of length $n$ over an alphabet of 
$q$ elements and $B_i, 0\le i\le n,$ its Hamming distance distribution
given by
%% Fix \bc' in the next sum!
\(
B_i={1\over |D|}\sum_{\bc'\in D}|\{\bc\in D\mid \dist(\bc,\bc')=i\}|.
\)
Suppose that $D$ is used for transmission over a $q$-ary memoryless
symmetric channel in which each symbol is unchanged with probability
$1-p$ and replaced by another symbol with one and the same
probability ${p\over q-1}.$ Let $\bc\in D$ be a vector sent
over the channel. The decoder tests the received vector $\by$ for 
containment in the code; if the test fails, it detects an error.
Thus, the only case when the error is not detected occurs
when the received vector $\by\in D\setminus\bc$.
The probability to receive a vector $\by$ if the sent vector is $\bc$
equals 
$P(\by|\bc)=\big({p\over q-1}\big)^{\dist(\by,\bc)}
(1-p)^{n-\dist(\by,\bc)}.$
Then the probability of undetected error for the code $D$ equals
\begin{align}
P_{ue}(D,p)&={1\over |D|}\sum_{\bc\in D}\sum_{\by\in D\setminus\bc}
P(\by|\bc)={1\over |D|}\sum_{\bc\in D}\sum_{\by\in D\setminus\bc}
\big({p\over q-1}\big)^{\dist(\by,\bc)}
(1-p)^{n-\dist(\by,\bc)}\nonumber \\
=&
\sum_{i=1}^{n} B_i \Big({p \over q-1}\Big)^i (1-p)^{n-i}.
\label{eq:Pud-classical}
\end{align}
The theory of error detection for classical codes is surveyed
in \cite{KK}.

Observe that this (classical) concept consists of two parts, the
definition of the error event and a way, (\ref{eq:Pud-classical}), to
compute its probability.  In the quantum case both parts are not nearly
as obvious.  One can consider error detection with stabilizer codes;
this produces a definition similar to the above one. It is not
difficult to show that the probability of undetected error in this
case can be computed via the weight enumerators of quantum codes in a
way analogous to (\ref{eq:Pud-classical}).  This definition can be
generalized in two different ways.  First, it is desirable to extend
it to cover all quantum codes; second and more importantly, from the
physical point of view if the angle between the received vector and
the transmitted vector is very small, it is natural to assume that no
error has occurred since the measurement of the state of the system
produces almost the same state as the transmitted one.  On the other
hand, if the received vector is a valid code point but is orthogonal
to the transmitted one, then with probability one the error is not
detected.  Hence it is natural to define the undetected error event as
the average probability that the state received from the channel is
orthogonal to the transmitted one.

A further generalization, which is physically perhaps the most important,
involves transmission of completely entangled states and
studying the probability of preserving the original entanglement.

The goal of our paper in the first part is to prove that in all
the situations mentioned, 
one can introduce a consistent definition of the undetected error
event, and that the actual functional on the quantum code
accounting for this event is essentially the same. This enables
us in the second part \cite{ABKL} to prove that there exist quantum codes
with exponentially falling probability of undetected error and
to derive bounds on this exponent.

The first part of the paper is organized as follows.
In Section \ref{qcodes} we recall some notions of the theory
of quantum codes, most importantly, the weight enumerators.
In Section \ref{def} we define and compute the probability of undetected
error for stabilizer and unrestricted codes. The answers
differ by a constant factor; hence their dependence on the
code is the same in both cases.  In
Section \ref{section:composite} we study the event of undetected 
error in the case of completely entangled states and  
again arrive to the same functional as in the previous sections. 
Section \ref{section:enumerators} is devoted to the study of 
polynomial invariants of quantum codes. We show that ``unitary 
weight enumerators'' of \cite{ref 2} are known in classical coding 
theory under the name of binomial moments of the distance distribution. 
They were introduced by MacWilliams in \cite{mac63}. They were 
studied extensively in \cite{ash99} (see also related works referenced 
there), partly because they are convenient for bounding below the 
probability of undetected error.

\section{Quantum Error Correcting Codes}
\label{qcodes}
In this section we review basic facts of the theory of
quantum codes relevant to our study, focusing on the error
process in the depolarizing channel and weight
enumerators of codes.

We begin by setting up basic linear-algebraic notation 
and reminding the reader of elementary quantum-mechanical 
operations performed on state vectors.
General sources for the relevant aspects of quantum theory
are books \cite{holevo}, \cite{peres}; a treatment that
highlights 
the context of information transmission is given in \cite{preskill}.
Below by bold letters $(\bv, {\bf w}, \dots)$ we denote complex
column vectors. For a given vector $\bf v$ we denote by $\bv^\ast $
its conjugate transpose. Let ${\cal H}_n={\mathbb C}^{2^n}$ denote the complex
$2^n$-dimensional space. Let us fix an orthonormal basis in this
space; denote it by $\bv_1,\dots, \bv_{2^n}$. Observe that
once the basis is fixed, a basis vector can be referred to by its
number, $i$. This is employed when the states and operations
are written in the Dirac notation, used in much of the physics
literature on quantum codes. The correspondence is established as
follows:
\[
\bv_i \leftrightarrow |i\rangle, \quad
\bv_i^\ast  \leftrightarrow \langle i|, \quad
\bv_i^\ast  B \bv_i \leftrightarrow \langle i|B|i\rangle,
\]
where $B$ is a $2^n\times 2^n$ matrix.
This correspondence is discussed from the physcal perspecttive in
\cite{peres}; mathematically oriented readers might enjoy
the discussion in \cite{kos}.

Let $V$ and $W$ be two subspaces of ${\cal H}_n$ and $A$ and $B$
be linear operators on $V$ and $W$, respectively.
%% Why introduce this notation but leave the tensor product
%% for operators? Is it used in a significant way.
%% I'd say, put in the tensor product explicitly everywhere.
Consider the linear operator $C=A\otimes B$ on $V\otimes W$.
Let $\{\bv_1,\bv_2,\dots\}$ be an orthonormal basis of $V$ and
$\{\bw_1,\bw_2,\dots\}$ an orthonormal basis of $W$.
The {\em partial trace} of $C$ over $V$ by definition equals
$$
\tr_V(C):=\tr(A) B=\sum_i (\bv_i^\ast   A\bv_i) B.
$$
Here $\tr(A)=\tr_V(A)=\sum_i (\bv_i^\ast   A\bv_i)$ is the trace of $A$.
Similarly, one can define the partial trace over $W$ as follows:
$$
\tr_W(C):=A \tr(B)=\sum_i A  (\bw_i^\ast  B\bw_i),
$$
Obviously,
\begin{equation}\label{eq:partial-composition}
%% The righthand side is not well defined!
%% The domain of \tr_V does not contain the range
%% of \tr_W. You need to extend the definition of \tr_V
\tr_{V\otimes W}(C)={\tr}_V(\tr_W(C)).
\end{equation}
 
Any linear operator $C$ on $V\otimes W$ 
can be written in the form
$$
C=\sum_j A_j\ten B_j, 
$$
where $A_j$ and $B_j$ are operators on $V$ and $W$, respectively.  
This representation is generally not unique. However, it is not 
difficult to check that the partial trace
\begin{equation} \label{partial_tr}
\tr_V(C)=\sum_j \tr(A_j) B_j, \;\tr_W =\sum_j A_j \tr(B_j).
\end{equation}
is a well-defined function.
(In fact, it is possible to give an invariant definition
of the trace \cite[p.130]{prasolov}.)

Let us proceed to the definition of quantum codes and
operations.
A {\em qubit} is a two dimensional Hilbert space.
A state (more precisely, a pure state) 
of a qubit is a unit vector $\bv$ in the Hilbert space ${\mathbb C}^2$.
Physically, a qubit occurs as
a spin one-half particle, for example.
Qubits are combined to form larger systems by taking
the tensor products of the Hilbert spaces. The state of 
$n$ qubits is therefore described by a unit vector in 
$({\mathbb C}^2)^{\otimes n}$ ($n$th tensor power).
A {\em quantum code} $Q$ is a linear subspace of 
${\cal H}_n$. We denote a code of dimension $K$ by $Q((n,K)).$

{\em Remark.} There is a subtle point about this definition.
Namely, although $Q$ is a linear subspace, proportional
vectors account for one and the same state, so in effect
we only deal with vectors of unit norm. (Below we write
$\bu\pll\bv$ to indicate that the vectors $\bu$ and $\bv$ are collinear.) 
Therefore, $Q$ can be thought of as the projective space 
${\bf P}{\mathbb C}^{K-1}$.
However, defining $Q$ in this way poses problems for
tensoring it with other subspaces (systems). Therefore,
we prefer the definition given above.
On the other hand, averaging over the code, we integrate only
over the vectors in $Q$ of unit norm. This convention, adopted 
with some abuse of notation, is valid for the entire paper.

The number $R_Q={\log_2 K \over n}$ is called the {\em rate} of $Q$.
A code vector (a state of $n$ qubits) ``sent'' over a quantum channel
is subjected to an error process that can alter the amplitude and/or
the phase of some of the qubits.  At the receiving end the decoder
attempts to recover the state sent. The decoder is a quantum computing
device; so it can perform only unitary rotations and measurements.  Let us
proceed to describing them in more detail.

Let $\bv \in {\cal H}_n,\, ||\bv||=1.$
A {\em unitary rotation} $U$ of the state $\bv\in {\cal H}_n$ produces a
state $U\bv$. A {\em measurement} of $\bv$ is a probabilistic operation
performed with respect to a set of projections
$P_1, P_2, \ldots ,P_t$ that satisfy the properties
\begin{gather*}
P_iP_j = \delta_{ij}P_i,  \\
P_1+P_2+ \ldots +P_t=\mbox{id}.
\end{gather*}
Measuring $\bv$ with respect to $P_1, P_2, \ldots ,P_t$ results in 
projecting it by one of the operators $P_i,1\le i\le t$. The resulting
state has the form
\begin{equation}\label{eq:state}
{P_i\bv \over \sqrt{\bv^\ast P_i\bv}},
\end{equation}
for some $i$. The probability for the $i$th projection
is $\bv^{*}P_i\bv$. Upon performing the measurement
we obtain the state (\ref{eq:state}) and observe the number $i$.  

Suppose a code vector $\bv\in Q$ is sent through the
{\em depolarizing channel}. The outcome of the
transmission can be written as $E\bv$, where the error operator
$E$ has the form
%% It needs to be made clear how this acts on the space H_n.
\begin{equation}\label{E tensor}
 E=\tau_1 \otimes \tau_2 \otimes \ldots \otimes \tau_n,
\end{equation}
and each $\tau_i$ is either $I_2$ or one of the following (Pauli)
matrices:
\begin{equation}
\label{errorset}
\sigma_x=\left [
\begin{array}{cc}
0 & 1 \\
1 & 0
\end{array}
\right]\!,\;
\sigma_z=\left [
\begin{array}{cc}
1 & 0 \\
0 & -1 
\end{array}
\right]\!,\;
\sigma_y=\left [
\begin{array}{cc}
0 & -\remove{\iota} i  \\
\remove{\iota} i  & 0
\end{array}
\right]. 
\end{equation}
The {\em weight} of the error $E$ is the number of
nonidentity matrices in the expansion (\ref{E tensor}).
Matrices of the form (\ref{E tensor}) either 
commute or anticommute:
\[
E_1E_2=\pm E_2E_1.
\]
The choice of the sign is determined by either geometric \cite{CCKS}
or algebraic considerations \cite{Cal gf}; see (\ref{com}) below.
Note also that error operators are ``trace-orthogonal'':
\begin{equation}
\label{tr(e1e2)}
\mbox{Tr}(E_iE_j)=2^n\delta_{ij}
\end{equation}
 
Throughout the paper we assume that the channel is {\em symmetric},
i.e., the probability 
\begin{equation*}
\Pr(\tau_i)=\begin{cases}
1-p, &\tau_i=I_2\\
{p\over 3},&\tau_i\in\{\sigma_x,\sigma_y,\sigma_z\},
\end{cases}
\end{equation*}
where $0\le p\le 3/4.$ Different qubits are subjected to the
error process independently. 
Therefore we have
\begin{equation}\label{Bi}
\mbox{Pr}(E)=\Big({p\over 3}\Big)^{\wt(E)}(1-p)^{n-\wt(E)}.
\end{equation}

Similarly to classical codes one can introduce the {\em weight enumerator} 
$B(y)=\sum_{i=0}^n B_iy^i$ of a quantum  $((n,K))$ code \cite{ref 1}.
By definition, 
\begin{equation}\label{eq:B}
B_i:=\frac{1}{K^2}\sum_{wt(E)=i} \mbox{Tr}^2(EP).
\end{equation}
Here $P$ is the orthogonal projection on $Q$.
Another weight enumerator associated with $Q$ \cite{ref 1}
is given by $B^\bot(y)=\sum_{i=0}^n B^\bot_iy^i$, where 
\begin{equation}\label{eq:Bperp}
B^{\bot}_i:=\frac{1}{K}\sum_{wt(E)=i} \mbox{Tr}(EPEP).
\end{equation}
Some reflection shows that both $B_i$ and $B_i^\bot$ are real.
Indeed, by (\ref{errorset}) error operators $E$ are Hermitian. 
Of course, $P$ is also Hermitian, and so is $PEP.$ 
Eigenvalues of Hermitian operators are real; therefore,
both $\mbox{Tr}(EP)$ and $\mbox{Tr}(EPEP)$ are real.  

The following two theorems account for the role of the
weight enumerators in the theory of quantum codes.
\begin{thm}
\label{thm:MacW}{\rm \cite{ref 1}}
\begin{equation}\label{mc}
B(y)={1\over 2^n K}B(1+3y,1-y)
\end{equation}
\remove{\begin{equation}
\label{mc}
B_i=\frac{1}{2^nK}\sum_{j=0}^n B^{\bot}_j P_i(j),
\end{equation}
where for $0\le i\le m$ and any $x$, 
\(
P_i(x):=\sum_{j=0}^n (-1)^j 3^{i-j}{x \choose j}{n-x \choose i-j}.
\)}
\end{thm}
 
\begin{thm}{\rm \cite{ref 1}}
\label{thm:enum}
Let $Q$ be a quantum code with weight
distributions $B_i$ and $B^{\bot}_i$. Then

\begin{itemize}
\item[{\it i$)$}] $B_0=B^{\bot}_0=1$ and
$B^{\bot}_i\ge B_i\ge 0$  $(1\le i\le n)$
 
\item[{\it ii$)$}] the minimum distance of $Q$ equals
$t+1,$ where $t$ is the largest integer
such that $B_i=B^{\bot}_i, 0\le i\le t.$
\end{itemize}
\end{thm}

\remove{ 
Taking into account that $B_0=B^{\bot}_0=1, P_0(i)=1,$   and  (\ref{mc}), we 
get 
\begin{equation}
\label{S}
K=\frac{1}{2^n}\sum_{i=0}^n B_i^{\bot}.
\end{equation}
}

These theorems are readily seen to be analogous to the MacWilliams
transform and surrounding results in classical coding theory.
A particular case of classical codes is formed by additive (or group)
codes \cite[p. 84]{del73}. An additive code $C$ is a subgroup of
the additive group of ${\bf Z}_q$. 
For such codes it is possible to define a dual code
$C^\bot$, and the Hamming weight enumerators of $C$ and $C^\bot$
are connected by the MacWilliams equation (\ref{mc}).
A similar concept, {stabilizer codes}, was introduced in
quantum coding theory in \cite{Cal geom},\,\cite{Gottesman}.

To define stabilizer codes, let us provide more details on error
operators (\ref{E tensor}). A natural framework to study them
is that of orthogonal geometry and the theory of finite groups
\cite{CCKS}. Matrices of the form $\alpha E,$ where $\alpha^2=\pm 1$ 
and $E$ is given by (\ref{E tensor}), form a group,
say ${\cal E}_n$, isomorphic to an extraspecial
$2$-group of order $2^{2n+2}$ \cite{CCKS}. Its center
is $Z=\{\pm I_{2^n},\pm i I_{2^n}\}$. Let
$\bar{\cal E}_n={\cal E}_n/Z$ be the quotient group, formed
by matrices of the form (\ref{E tensor}) with constant factors disregarded.
In what follows an error operator $E$ can be either from ${\cal E}_n$ or 
$\bar{\cal E}_n$. 
If $E\in \bar{\cal E}_n$ then we will assume that $E$ is of the form (\ref{E 
tensor}), that is we take $E$ as a coset representative.

{\em Definition.} A quantum code $Q$ is called a {\em stabilizer} code
if there exists an Abelian subgroup ${S}$ of the group 
${\cal E}_n$ such that $Q$ is an eigenspace of $S$ with eigenvalue $1$. 
In other words, a code $Q$ is a stabilizer code if
\[
Q=\{\bv\in {\cal H}_n\mid \forall_{E\in S}\, E\bv=\bv\}. 
\]
If the order of $S$ is $2^{n-k}$, then  $\dim(Q)=2^{k}$
\cite{Cal gf},\, \cite{Gottesman},
and so the dimension of a stabilizer code always equals
an integer power of $2$. A stabilizer code of length $n$ and
dimension $2^k$ is denoted by $Q[[n,k]].$

Let $F:={\mbox GF}(4)=\{0,1,\omega,\omega^2\}.$
In the remaining part of this section
we explain a way to associate to a quantum stabilizer code two 
additive codes $C$ and $C^\bot$ over $F$ in such a manner that 
$B(x,y)$ and $B^\bot(x,y)$ are their respective weight enumerators.
This connection is the main discovery of \cite{Cal gf}.
It is also of key importance for both parts of our study.

We begin by establishing a bijection $\bar\phi$
between the set of matrices $\{I_2,\sigma_x,\sigma_z,\sigma_y\}$
and the elements of $F $ as follows:
$$
\bar\phi(I_2)=0,\; \bar\phi(\sigma_x)=1,\;  \bar\phi(\sigma_z)=
\omega,\; \bar\phi( \sigma_y)= \omega^2,
$$
where $\omega$ is a primitive element of $F $. 
This bijection naturally extends 
to a bijection $\phi_n$ between $\bar{\cal E}_n$ and $F^n$
and defines a mapping $\phi_n:{\cal E}_n\to F^n$ as a composition
of $\bar\phi_n$ and the factorization mapping.
For $a\in F $ let $\bar a$ be its conjugate under the action of 
the Galois group. In particular, this action transposes 
$\omega$ and $\bar\omega=\omega^2.$
Let $\ast$ be the inner product in $F^n$
defined by
\begin{equation}
\label{inpr}
{\bf u}\ast \bv=\mbox{tr}\big( \sum_i v_i 
\bar u_i\big)=\sum_{i}(v_iu_i^2+v_i^2u_i), 
\end{equation}
where $\mbox{tr}(x):=x+x^2$ is the trace from $F $ to $GF(2).$
Key properties of error operators from ${\cal E}_n$ in this context 
are the following \cite{Cal geom}, \cite{Cal gf}:
\begin{equation}
\label{com}
E_1E_2=\begin{cases}E_2E_1 &\be_1\ast \be_2=0,\\
-E_2 E_1 &\be_1\ast\be_2\ne 0\end{cases}
\end{equation}
where ${\be}_i=\phi_n({E}_i)$,
and
%% The following is not iff, unless the E_i are projected
%% to the abelian group first.
%% Also, make sure to avoid confusion on where S lives below.
\[
{E}_1 {E}_2={E}_3\;\Rightarrow\;\be_1+\be_2=\be_3.
\]
Thus, $\bar\phi_n$ establishes a group isomorphism between $\bar{\cal E}_n$
and the additive group of $F^n$; so $\bar{\cal E}_n$ is elementary
Abelian of order $2^{2n},$ and $\phi_n$ is a homomorphism.
Another useful property of error operators is
\begin{equation}
\label{tr(e1e2e1e2)}
\mbox{Tr}(E_1E_2E_1E_2)=(-1)^{e_1\ast e_2} 
\mbox{Tr}(E_1E_2E_2E_1)=(-1)^{e_1\ast 
e_2}2^n.
\end{equation}

Now let $S<{{\cal E}}_n$ be a subgroup. Let
\[
C:=\{\bc\in F^n\mid \bc=\phi_n(E) \mbox{ for some }E\in S\}.
\]
Obviously, $C$ is an
additive subgroup in the group $F^n$. Let
\[
C^\bot=\{\bc\in F^n\mid \forall_{\bc'\in C}(\bc\ast\bc')=0.\}
\]
Letting $\bc$ range over the entire 
$C^\bot$ defines a subgroup 
$\bar{S}^\bot=\{\bar\phi^{-1}_n(\bc)\}<\bar{\cal E}_n$. 
By definitions of $C, C^\bot,$ and (\ref{com}) 
we see that any matrices $E\in S,\,E'\in S^\bot$ commute.
%% The following note is wrong under S < \bar E_n.
Note also that $S$ is Abelian; by (\ref{com}) this implies
that $\bc'\ast\bc''=0$ for any vectors $\bc',\bc''\in C$;
hence
\[
C\subseteq C^\bot
\]
and therefore, $S< S^\bot$ and $\bar{S}< \bar{S}^\bot$.

The following theorem completes our task.
\begin{thm}\label{thm:C-C}{\rm \cite{ref 12}}
Let $Q$ be a stabilizer code, $B(x,y)$ and $B^\bot(x,y)$
its weight enumerators, and let $C$ and $C^\bot$ be the codes
over $F^n$ defined above. Then the Hamming weight enumerator of $C$ 
$($resp., $C^\bot)$ is $B(x,y)$  $(B^\bot(x,y)).$
\end{thm} 
The proof relies on the following lemma whose short proof we
include for completeness.
\begin{lemma}{\rm\cite{ref 12}}  The orthogonal 
projection $P$ on $Q$ can be written in the following form:
\begin{equation}
\label{proj}
P={1 \over 2^{n-k}} \sum_{E\in S} E.
\end{equation}

\end{lemma}
\proof
A linear operator is a projection if and only if it is
idempotent. A projection is orthogonal if and only it is self-adjoint. 
Self-adjointness of $P$ follows from (\ref{errorset}) and (\ref{E
tensor}). $P$ is also idempotent.
$$
P^2=\Big({1 \over 2^{n-k}}\Big)^2\sum_{E'\in S} E' 
\sum_{E\in S} E =\Big({1 \over 2^{n-k}}\Big)^2\sum_{E\in S} 2^{n-k}E=P.
$$
Let us compute the dimension of the space $\{P\bv\mid\bv\in {\cal
H}\}$. From (\ref{E tensor}) we see that $\tr(E)=0$ unless
$E=I_{2^n}.$ Hence $\tr (P)=2^k=\dim Q.$ Finally, for any 
$\bv\in Q$ we have $E\bv=\bv$ by definition of $Q$; hence also
$P\bv=\bv.$
\QED

The proof of Theorem (\ref{thm:C-C}) is now accomplished by
invoking (\ref{tr(e1e2e1e2)}) and (\ref{tr(e1e2)}).

\remove{
To prove the theorem, invoke (\ref{tr(e1e2e1e2)}) and (\ref{tr(e1e2)})
to obtain
\begin{equation}
\label{stbi}
\mbox{Tr}(E'P)={1 \over 2^{n-k}} \sum_{E\in S}\mbox{Tr}  (E'E)= 
2^k\chi\{E'\in S\},
\remove{\left\{\begin{array}{lc}
2^k, & E'\in S,\\
0, & else,
\end{array}\right.} 
\end{equation}
\begin{align}
\mbox{Tr}(E'PE'P)&={1 \over 2^{2(n-k)}}
\sum_{E\in S}\mbox{Tr}(E'EE'E)\nonumber \\
&={2^n \over 2^{2(n-k)} }  \sum_{E\in S} (-1)^{\be'\ast \be}=
2^k\chi\{E'\in S^\bot\},
\remove{
\left\{ \begin{array}{lc}{lc}
2^{k}, & E'\in S^\bot,\\
0, & else,
\end{array}\right.} \label{stbiperp}
\end{align}
where $\chi$ is the indicator function.
Together with (\ref{eq:B}) and (\ref{eq:Bperp}) this implies the theorem. 
\QED
}

\section{Quantum undetected error} 
\label{def}

In this section we study error detection with quantum
codes used over the depolarizing channel.
We examine a number of possible definitions of undetected
error and establish their equivalence.

For a quantum code $Q$ we denote by $Q^\bot$ its orthogonal code
under the standard Hermitian inner product on ${\cal H}_n$.
The orthogonal projection on $Q^\bot$ is denoted by $P^\bot$.

Let us first describe the {\em quantum error 
detection protocol.} The transmitted vector $\bv\in Q$ is 
corrupted in the channel by the action of an error operator 
$E\in \bar{\cal E}_n$. The received vector has the form $\bw=E\bv$.
In order to detect the error we perform the measurement of 
$\bf{w}$ with respect to the projections $(P, P^\bot)$. If the result 
of this measurement is contained $Q^\bot$ we detect 
an error. Otherwise we assume that there is no error and that 
the  code vector $P\bw$ was transmitted. 

\nd{\sc a. stabilizer codes} Let us begin with the easier case
of stabilizer codes. Suppose $S$ is a stabilizer group of a code $Q$.
Below by $C$ we denote the code obtained
by evaluating $\phi_n(E)$ on all $E\in S$.

Let $E\in \bar{\cal E}_n$ and $\bv\in Q$. Suppose $\bz$ is
a result of the measurement of $\bw=E\bv$  with respect to the 
system $(P,P^\bot)$. Let us 
introduce the indicator function $\delta(E,\bv)$ as follows: 
\begin{equation*}
\delta(E,\bv)=\begin{cases}
1 &\bz\in Q\mbox{ and }\bz \not\pll \bv\\
0 &\mbox{otherwise}.
\end{cases}
\end{equation*}
Note that if $\bw\in Q$ then measuring it 
with respect to the system $(P,P^\bot)$ with 
probability 1 we obtain a code vector. As shown below, for stabilizer
codes the vector $\bw$ is contained either in $Q$ or in $Q^\bot;$
hence in this case $\delta$ is deterministic.
%% might want to be more explicit: a function of E and v.

{\em Definition.} Let $Q$ be a stabilizer code. Its probability
of undetected error in a  depolarizing channel with
error probability $p$ is defined as
\begin{equation}
\label{p2}
P_{ue}^{(s)}(Q,p)=\sum_{E\in \bar{\cal E}_n} \mbox{Pr}(E)\int_{\bv\in 
Q}\delta(E,\bv) 
d\nu(\bv),
\end{equation}
where $\nu(\cdot)$ is the normalized uniform measure on $Q$. 

To compute the probability of undetected error let us first specify
the error event, i.e., examine the case $\delta(E,\bv)=1$ in more detail.
Consider a possible effect of 
error operators on vectors from $Q$. For an
error operator $E$ let $\be=\bar\phi_n(E).$ 
There are the following three possibilities. 

1) If $E\in \bar{S}$ then $E\bv \pll\bv$ for any $\bv \in Q$; so this
error has no effect on the transmission. 

2) Let  $E\not\in \bar{S}^\bot$. This means that $\be\not\in C^\bot$ 
and so $\be\ast \be'\not =0$ for some $\be'\in C$. Hence
$EE'=-E'E$ and $E'E\bv=-EE'\bv=-E\bv$. Thus $E\bv$
is contained in the eigenspace of $E'$ with eigenvalue  $-1$. 
Since $E$ is a Hermitian operator, its eigenspaces are pairwise
orthogonal. 
Thus
$$
{\bf w^\ast }E\bv=0
$$
for any ${\bf w}\in Q$. 
In other words $\bw \in Q^\bot$, and making measurement 
with respect to $(P, P^\bot)$ we will detect the presence of
error with probability $1$. 

3) Now let $E\in \bar{S}^\bot \setminus \bar{S}$, i.e., $\be\in C^\bot \setminus 
C$. 
Then  $\be\ast \be'=0$ for all $\be'\in C$. Hence  by (\ref{com}) 
$EE'=E'E$ for all $E'\in S$, and so
$E'E\bv=EE'\bv$. Hence $E\bv$ is an eigenvector of $E'$ with 
eigenvalue $1$ and therefore $E\bv\in Q$. However, $E\not\in \bar{S};$ so
$E\bv\not\pll\bv$ for some $\bv\in Q.$
Thus, $E$ acts on $Q$ by rotation, possibly with some invariant
directions.  

Note that in contrast with classical linear codes there
are errors whose detection depends on the transmitted
vector. 

The above information enables us to compute the probability
of undetected error.
\begin{prop}
\begin{align}
P_{ue}^{(s)}(Q,p)&=\sum_{E\in \bar{S}^\perp\setminus \bar{S}} \Pr(E)
\label{eq:stabilizer1}\\
&=\sum_{i=0}^{n} (B^\bot_i -B_i)  \left({p \over 3}\right)^i 
(1-p)^{n-i}\label{eq:stabilizer2}
\end{align}
\end{prop}
\Proof The proof follows the standard
proof of (\ref{eq:Pud-classical}),
relying upon (\ref{Bi}) and the definition of the weight
distributions $\{B_i\}$ and $\{B^\bot_i\}.$

Suppose $E\in \bar{S}^\bot\setminus \bar{S}$ and $\be=\bar\phi_n(E).$
Let 
\[
\widehat{Q}(E):=\{\bv\in Q\mid E\bv=\bv\}.
\]
Let us construct the code $C_{\be}$ by adjoining $\be$ to the basis
of $C$. Since $\be\ast\be=0$ for any $\be \in {\mathbb C}^{2^n}$, 
the new code is self-orthogonal.
Hence we can construct an $[[n,k-1]]$ quantum stabilizer code $Q(E)$ 
associated with $C_\be$. Obviously $Q(E)\subset Q$ and any $\bv\in Q(E)$ 
is stabilized by $E$. It also follows from the definition of stabilizer codes
that all vectors of  ${\mathbb C}^{2^n}$ stabilized by both $S$ and $E$ belong 
to $Q(E)$. Hence 
$\widehat{Q}(E)=Q(E)$. 
Therefore, the size of $\widehat{Q}(E)$ does 
not depend on $E$, and for every $E$ the set $\widehat{Q}(E)$
is a $K-1$-dimensional subspace.  
Therefore $\nu(\widehat Q(E))=0.$ This implies our claim since
\begin{align*}
P_{ue}^{(n)}(Q,p)& = \sum_{E} \Pr(E)\int_{\bv\in Q}\delta(E,\bv) d\nu(\bv)\\
&=\sum_{E\in \bar{S}^\bot\setminus \bar{S}}\Pr(E)
\nu( Q\setminus\widehat{Q}(E))=\sum_{E\in \bar{S}^\bot\setminus 
\bar{S}}\Pr(E)\nu( Q)\\
&=\sum_{E\in \bar{S}^\bot\setminus \bar{S}} \mbox{Pr}(E).
\end{align*}
This proves (\ref{eq:stabilizer1}). To prove the final part,
note that $|\{E: \wt(E)=i, E\in \bar S^\bot\setminus \bar S\}|=B_i^\bot-B_i.$
\qed

This proposition shows that
if, speaking loosely, we assume that the error $E$ is not
detectable whenever $E\in \bar{S}^\bot \setminus \bar{S}$, ignoring vectors
from $\hat Q$, the probability of undetected error
will be still given by (\ref{eq:stabilizer1})-(\ref{eq:stabilizer2}).
This might motivate another definition of undetected error for
stabilizer codes, namely,
assuming the uniform distribution on $Q$, let us call an error $E$
undetectable if $E\in \bar{S}^\bot \setminus \bar{S}$.

\bigskip\nd{\sc b. nonstabilizer codes}.
In this part we consider a more general situation of $Q$
an arbitrary quantum code. Moreover, we take a viewpoint more
natural in the physical perspective, namely if a ``received'' vector
is close to the transmitted vector then we assume that little error
has occurred despite the fact that formally they are not equal.
This assumption is justified since a physical measurement 
in this case will not exhibit any difference between the
transmitted and the received vectors. 

More specifically, let $\bv\in Q$ and $\bw=E \bv$ for some error
operator $E$. If as a result of the measurement with respect to
$(P,P^\bot)$, $\bw$ is projected on $Q^\bot$, the error is detected.
Let us examine more closely the situation of $\bw$ projected on $Q$.
Let 
\begin{equation}\label{eq:z}
\bz={P\bw \over \sqrt{\bw^\ast  P\bw}}={PE\bv\over \sqrt{\bv^\ast  EPE \bv}}
\end{equation}
be this projection, which occurs with probability $\bv^\ast  EPE \bv$.
A natural proximity measure of $\bz$ and $\bv$ is the absolute
value of the angle $\angle(\bz,\bv)$
between them. If $\bz$ is sufficiently close 
to $\bv$, we assume that no error has occurred. 
By way of {thought experiment} suppose that for a given code
vector $\bv$ we measure
$\bz$ with respect to the system $(\bv \bv^\ast , P-\bv \bv^\ast )$.
Denote the result of this measurement by $\by_\bz$.
In other words, we represent $Q$ as a direct sum of $\bv$ and
a $K-1$-dimensional orthogonal subspace $Q_\bv$ and create a pair 
of projections, on the line given by $\bv$ and on $Q_\bv$.
The probability that after this measurement
$\bz\in Q$ projects on $\bv$ equals
$\bz^\ast \bv \bv^\ast \bz=cos^2 \angle(\bz,\bv);$
the probability of the complementary event is 
$\bz^\ast (P-\bv\bv^\ast )\bz$. In the former case
we assume that the error has no effect on transmission; in the
latter that the error is undetectable.

The overall probability that if $\bv\in Q$ is subjected to
an error operator $E$, then $\by_\bz$ is a code vector orthogonal 
to $\bv$ equals $||(I-\bv\bv^\ast )PE\bv||^2.$ Indeed, it equals
\begin{align*}
\Pr\{\bz\in Q\}\Pr\{\by_\bz\perp\bv|\bz\in Q\}&=
(\bv^\ast  EPE \bv)(\bz^\ast (P-\bv\bv^\ast )\bz )\\
&\stackrel{\mbox{\footnotesize(a)}}{=}
\bv^\ast  EP(I_{2^n}-\bv\bv^\ast )PE\bv\\
&\stackrel{\mbox{\footnotesize(b)}}{=}
\bv^\ast  EP(I_{2^n}-\bv\bv^\ast )(I_{2^n}-\bv\bv^\ast )PE\bv\\
&=||(I-\bv\bv^\ast )PE\bv||^2,
\end{align*}
where (a) follows upon substituting $\bz$ from (\ref{eq:z})
and recalling that $P$ is a projection on $Q$
and (b) uses the fact that $I_{2^n}-\bv\bv^\ast $ is a projection.
Concluding, we arrive at the following general definition of
the probability of undetected error.

\medskip
{\em Definition.} Let $Q$ be a quantum code used over a 
depolarizing channel with error probability $p$.
Then 
\begin{equation}\label{eq:def-non1}
P_{ue}^{(n)}(Q,p)=\sum_{E}\mbox{Pr}(E)\int_{\bv\in Q} 
||(I-\bv\bv^\ast )PE\bv||^2 d\nu(\bv),
\end{equation}
where $\nu$ is a normalized uniform measure on $Q$.
\medskip

This definition is more general than the one given for
stabilizer codes.

The main result that we prove regarding $P_{ue}^{(n)}(Q,p)$
is given in the following theorem, which shows that
$P_{ue}^{(n)}(Q,p)$ differs from $P_{ue}^{(s)}(Q,p)$ only
by a constant factor. 
\begin{thm}\label{thm:pud-non1}
Let $Q$ be an $((n,K))$ quantum code with
weight distributions $B_i$ and $B_i^\bot$, $0\le i\le n$.
Then
\begin{equation}\label{eq:pud-non1}
P_{ue}^{(n)}(Q,p)={K\over K+1}
\sum_{i=0}^n(B_i^\bot -B_i)\Big({p \over 3}\Big)^i 
(1-p)^{n-i}.
\end{equation}
\end{thm}
\Proof
The proof is accomplished by computing the integral in
(\ref{eq:def-non1}). It is clear that
\begin{align}
&\int_{\bv\in Q} ||(I-\bv\bv^\ast )PE\bv||^2d\nu(\bv) \nonumber \\
&=\int_{\bv\in Q} \bv^\ast  
EP(I-\bv\bv^\ast )(I-\bv\bv^\ast )PE\bv d\nu(\bv)\nonumber \\
&= \int_{\bv\in Q}  \bv^\ast  EPE\bv d\nu(\bv) - \int_{\bv\in Q} 
(\bv^\ast  
EP\bv)(\bv^\ast PE\bv) d\nu(\bv).\label{integrals}
\end{align}
To compute the integrals in (\ref{integrals})
we need the following lemmas proved in the appendix.

\begin{lemma}
\label{lemma:int}
Let $Q$ be an $(n,K)$ quantum code. Let $P$ be the orthogonal 
projection on $Q$, and let $\mu(\bv)$ be the normalized uniform 
measure on $Q$. Then
$$
\int_{\bv\in Q} \bv\bv^\ast  d\mu(\bv)={P \over K}
$$
\end{lemma}

\begin{lemma}
\label{vv_vv}\remove{
Let $\bv_i$ be an orthonormal basis of ${\mathbb C}^K$. Then} 
\begin{equation}
\label{int_3}
\int_{\bv\in Q} \bv\bv^\ast \ten \bv\bv^\ast  
d\mu(\bv)={1\over K(K+1)} 
\left(\sum_{i,j}\bv_i\bv_i^\ast \ten \bv_j\bv_j^\ast +
\bv_i\bv_j^\ast \ten 
\bv_j\bv_i^\ast \right)
\end{equation}
\end{lemma}

Using these lemmas, let us compute the integrals in (\ref{integrals}).
By Lemma \ref{lemma:int} we have for the first term
\begin{align}
\int_{\bv\in Q} \bv^\ast EPE\bv d \mu(\bv)
&\stackrel{\mbox{\footnotesize (a)}}{=} \int_{\bv\in Q}
\tr(EPE\bv\bv^\ast ) 
d\mu(\bv)\nonumber \\
&=\tr\Big (EPE\int_{\bv\in Q} \bv\bv^\ast d \mu(\bv)\Big) \nonumber \\
&={1\over K} \tr(EPEP),\label{int_1}
\end{align}
where (a) is obtained by replacing a scalar by its trace.

Let us compute the second term in (\ref{integrals}).
The code $Q$ is a $K$-dimensional linear space; so it
is isomorphic to the complex space ${\mathbb C}^{K}$. The following
calculations are simplified by performing them on ${\mathbb C}^{K}$ 
instead of $Q.$ 

Note that for any operators 
\begin{equation}\label{eq:trace-cyclic}
\tr(ABC)=\tr(BCA).
\end{equation}
Therefore, we can rewrite our integral as
\begin{equation}
\label{int_2}
\int_{\bv\in {\mathbb C}^K} \bv^\ast EP\bv\bv^\ast PE\bv d\mu(\bv)
=\int_{\bv\in {\mathbb C}^K} 
\tr(EP\bv\bv^\ast PE\bv\bv^\ast ) d\mu(\bv).
\end{equation}

Let $L_n$ be the space of linear operators on ${\cal H}_n$.
Consider the bilinear functional
\begin{align*}
\Phi_1:\; &L_n\times L_n \to {\mathbb C}\\
&(A,B)\mapsto \tr(CAC^\ast B).
\end{align*}
By the definition of the tensor product \cite{kos}, there exists
a (universal) bilinear map $t:\,L_n\times L_n\to L_n\ten L_n$
such that $\Phi_1=\Phi_2\circ t$, where $\Phi_2$ is a {\em linear}
functional defined by
\begin{align*}
\Phi_2:\; &L_n\otimes L_n \to {\cal H}_n\\
&A\ten B\mapsto \tr(CAC^\ast B).
\end{align*}

Now we have
\begin{align}
\int_{\bv\in {\mathbb C}^K} \bv^\ast EP\bv\bv^\ast  & PE\bv d\mu(\bv)
\stackrel{\mbox{\footnotesize(a)}}{=}\int_{\bv\in {\mathbb C}^K} 
\Phi_2(\bv\bv^\ast \ten \bv\bv^\ast ) d \mu(\bv)  \nonumber\\
&\stackrel{\mbox{\footnotesize(b)}}{=}\Phi_2\Big(\int_{\bv\in 
{\mathbb C}^K} \bv\bv^\ast \ten \bv\bv^\ast  d 
\mu(\bv)\Big)\nonumber\\
&\stackrel{\mbox{\footnotesize(c)}}{=}{1 \over K(K+1)} \Phi_2
\Big(\sum_{i,j=1}^K
\bv_i\bv^\ast _i\ten \bv_j\bv_j^\ast +\bv_i\bv_j^\ast \ten \bv_j
\bv_i^\ast \Big)\nonumber \\
&\stackrel{\mbox{\footnotesize(d)}}{=}{1\over K(K+1)}\Big[\sum_i 
\tr(EP\bv_i\bv_i^\ast PE\sum_j 
\bv_j\bv_j^\ast )+\sum_{i,j}\tr(EP\bv_i\bv_j^\ast PE\bv_j\bv_i^\ast )
\Big]\nonumber \\
&\stackrel{\mbox{\footnotesize(e)}}{=}{1\over K(K+1)}\Big[\sum_i 
\tr(EP\bv_i\bv_i^\ast PEP)+\sum_{i,j}\tr(\bv_i^\ast EP\bv_i)
\tr(\bv_j^\ast PE\bv_j)\Big]\nonumber \\
&\stackrel{\mbox{\footnotesize(f)}}{=}{1\over K(K+1)}
\Big[\tr(PEPEP\sum_i \bv_i\bv_i^\ast )+
\tr(EP\sum_i\bv_i\bv_i^\ast )\tr(PE\sum_j\bv_j\bv_j^\ast )\Big]\nonumber\\
&\stackrel{\mbox{\footnotesize(g)}}{=}{1\over K(K+1)}
\Big[\tr(PEPE)+\tr(EP)^2\Big]\label{int_4} 
\end{align}
Here (a) follows by (\ref{int_2}) and the definition of $\Phi_2,$
(b) holds true by linearity of $\Phi_2,$ (c) follows by Lemma \ref{vv_vv},
(d) is by definition of $\Phi_2$, (e) uses the fact that 
$P=\sum_i \bv_i\bv_i^\ast,$ (f) uses (\ref{eq:trace-cyclic}) and the
additivity of trace, and (g) follows by (\ref{eq:trace-cyclic}) and
the fact that $P$ is idempotent.

Substitution of (\ref{int_1}) and (\ref{int_4}) in (\ref{integrals}) 
gives
$$
\int_{\bv\in Q} ||(I-\bv\bv^\ast )PE\bv||^2d\nu(\bv)=
{K\over K+1}\Big[{1\over 
K}\tr(EPEP)-{1\over K^2}\tr(EP)^2\Big];
$$
by (\ref{eq:B})-(\ref{eq:Bperp}) this proves the theorem. \qed

\remove{
and eventually
$$
P_{ue}^{''}(Q,p)={K\over K+1}\sum_{i=0}^n(B_i^\bot -B_i)\left({p \over 
3}\right)^i (1-p)^{n-i}
={K\over K+1}P_{ue}(Q,p).
$$}

\section{Composite systems}
\label{section:composite}

In this section we study the most general problem setting for
quantum error detection. 
The general idea is to take into account not only the error
process but also the relationship of the current code vector
with states of other quantum systems.
More specifically, the qubits of the current vector
can be entangled with other qubits that may not even take
part in the transmission. We would like to study error 
detection that takes into account not only the error process 
but also this entanglement. The overall goal is to evaluate
how well the original entanglement is preserved under the
action of errors. Somewhat surprisingly, though this 
definition of undetected error is more broad, the actual
functional is again the same as studied above. 

This problem has no direct analogy in classical information 
transmission where typically one can
study the effect of the error process on the transmitted
vector without considering the influence on it of other
parts of the message.

Let $Q$ be a quantum code that is a part of a combined
system $QR=Q\ten R,$
where $R$ is another $K$-dimensional subspace.
A generic element of $QR$ can be written as $\sum_{i=1}^K \bv_i \ten
\bw_i,$ where $\bv_i$ and $\bw_i$ are basis vectors of space $Q$ and 
$R$ respectively.
Upon normalization we obtain a {\em completely entangled
state} of the composite system,
$$
\bb_{QR}={1 \over \sqrt{K}} \sum_{i=1}^K \bv_i \ten \bw_i.
$$
Let us assume that we transmit or store in quantum memory
only qubits of $Q$. These qubits are subjected to the
error process described above; so qubits of $R$ remain 
error-free. The ``received'' state has the form
\[
\widehat{\bb}_{QR}={1 \over \sqrt{K}} 
\sum_{i=1}^K (E_Q\ten I_R) \bv_i \ten 
\bw_i,
\]
where $E_Q$ is an error operator on $Q$ and $I_R$ is the 
identity operator on $R$.
At the ``receiving end'' we again apply the same decoding, namely, 
measure the state with respect
to $(P_Q,P_Q^\bot).$
As above, the error is not detected if after this measurement
we obtain a code vector that is orthogonal to the transmitted
vector $\bv$.
Therefore, in analogy with (\ref{eq:def-non1})
let us define the probability of undetected error
for the case of composite systems as follows
\begin{equation}\label{eq:def-combined}
P^{(c)}_{ue}(Q,p):=
\sum_{E_Q} \mbox{Pr}(E_Q)||(I_R\ten I_R-\bb_{QR}\bb_{QR}^\ast )
(P_Q\ten I_R)(E_Q\ten I_R)\bb_{QR}||^2 
\end{equation}
The main result of this section is given by the following theorem.
\begin{thm}\label{thm:combined}
Suppose we transmit completely entangled states $\bb_{QR}$
over a  depolarizing channel with error probability
$p$. Then
\[
P^{(c)}_{ue}(Q,p)=\sum_{i=0}^n (B_i^\bot-B_i)  \left({p \over 3}\right)^i 
(1-p)^{n-i},
\]
where $B_i, B_i^\bot$ are the weight enumerators of the code $Q$.
\end{thm}
\Proof
Let us compute the summation term in (\ref{eq:def-combined}). We have
\begin{align}
||(I_R\ten I_R-&\bb_{QR}\bb_{QR}^\ast )
(P_Q\ten I_R)(E_Q\ten I_R)\bb_{QR}||^2\nonumber\\
&=\bb_{QR}^\ast \Big[ (E_Q\ten I_R)(P_Q\ten I_R) (I_Q\ten I_R-b_{QR} 
\bb_{QR}^\ast) (I_Q\ten I_R-\bb_{QR}\bb_{QR}^\ast) \nonumber\\
&\qquad\times(I_Q\ten I_R-\bb_{QR}\bb_{QR}^\ast)  
(P_Q\ten I_R)(E_Q\ten I_R) \Big] \bb_{QR}\nonumber\\
&=
\bb_{QR}^\ast (E_Q\ten I_R)(P_Q\ten I_R)(E_Q\ten I_R) 
\bb_{QR} \nonumber \\
&\quad-(\bb_{QR}^\ast (E_Q\ten I_R)(P_Q\ten I_R) 
\bb_{QR})( \bb_{QR}^\ast (P_Q\ten I_R) (E_Q\ten I_R) \bb_{QR}).
\label{cos}
\end{align}
The first term in this expression can be computed as follows 
\begin{align}
 \bb_{QR}^\ast (E_Q\ten I_R)&(P_Q\ten I_R)(E_Q\ten I_R) \bb_{QR} \nonumber \\
& \stackrel{\mbox{\footnotesize(a)}}{=}  
\tr_{QR}( (E_Q\ten I_R)(P_Q\ten I_R)(E_Q\ten I_R) 
\bb_{QR} \bb_{QR}^\ast)  \nonumber  \\
&\stackrel{\mbox{\footnotesize(b)}}{=}    
\tr_Q\Big(\tr_R\,\big( (E_Q\ten I_R)(P_Q\ten I_R)(E_Q\ten I_R) \bb_{QR} 
\bb_{QR}^\ast\big)\Big)  \nonumber  \\
&\stackrel{\mbox{\footnotesize(c)}}{=}    
\tr_Q\biggl(\tr_R\,\Big({1\over K} \sum_{i,j} 
E_Q P_Q E_Q \bv_i \bv_j^\ast \ten 
\bw_i \bw_j^\ast \Big)\biggr)  \nonumber  \\
&\stackrel{\mbox{\footnotesize(d)}}{=}   
 {1 \over K} \tr_Q\Big(\sum_{i,j,l} E_QP_QE_Q \bv_i \bv_j^\ast\ten 
\bw_l^\ast \bw_i \bw_j^\ast \bw_l\Big)\\
&\stackrel{\mbox{\footnotesize(e)}}{=}  
{1 \over K} \tr_Q\Big(E_QP_QE_Q\sum_{l}\bv_l\bv_l^\ast\Big) \nonumber \\
&= 
{1\over K} \tr_Q\left(E_QP_QE_QP_Q\right),\label{eq:first_term}
\end{align}
where (a) is obtained upon replacing a scalar by
its trace and using (\ref{eq:trace-cyclic}),
(b) follows by property (\ref{eq:partial-composition})
of partial traces,
in (c) we substitute the definition of $\bb_{QR}$,
in (d) we compute the trace over $R$ and use
(\ref{eq:trace-cyclic}), and in (e) we convolve over
the dumb indices $i,j.$

Let us compute the second term in (\ref{cos}).
Proceeding as above, we obtain
\begin{align}
(\bb_{QR}^\ast &(E_Q\ten I_R)(P_Q\ten I_R) 
\bb_{QR})( \bb_{QR}^\ast (P_Q\ten I_R) 
(E_Q\ten I_R) \bb_{QR})  \nonumber  \\
&= \tr_{QR}( (E_Q\ten I_R)(P_Q\ten I_R)
\bb_{QR}\bb_{QR}^\ast) \nonumber\\
&\qquad\times\tr_{QR}[(P_Q\ten 
I_R)
(E_Q\ten I_R)\bb_{QR}\bb_{QR}^\ast]  \nonumber  \\
&={1\over K^2}   
\tr_Q(E_Q P_Q P_Q)\tr_Q(P_QE_QP_Q)\nonumber\\
&={1\over K^2} \tr_Q(P_QE_Q)^2.\label{eq:second_term}
\end{align}
Substitution of (\ref{eq:first_term}) and (\ref{eq:second_term}) 
in (\ref{cos}) together with (\ref{eq:B})-(\ref{eq:Bperp})
completes the proof.\qed

This concludes our main task for the first part of the paper.
We have proved that there exists a consistent definition
of the probability of undetected error for quantum codes
that can be given in the general case under natural
physical assumptions and in the case of stabilizer
codes analogously to the classical error detection.
The functional of undetected error on $Q$ in all the cases considered
is the same, up to a constant factor. 
Therefore, as in classical coding theory, we can study
performance of quantum codes under error detection.
The most important question in this context is to prove that
the probability of undetected error for the best possible
codes falls exponentially with code length $n$ and to exhibit
specific bounds on this exponent. Namely, let
\[
P_{ue}(n,K,p)=\min\limits_{Q\in {\cal H}_n\atop \dim(Q)=K} P_{ue}(Q,p).
\]
The second part of the paper is devoted to the study of this
function. We answer the main question in positive by proving the 
existence of families of stabilizer codes with exponential decline 
of $P_{ue}(K,n,p),$ and establish upper bounds on this function
for all quantum codes.

\section{Quantum weight enumerators}\label{section:enumerators}
In this section we focus on different forms of quantum
weight enumerators. We begin with a short
digression on classical enumerators. Let $D\subset {\mathbb F}_q^n$
be a linear code, $\supp(\bx)=\big\{e\in\{1,2,\dots,n\}| \bx_e\ne0\big\}$
the support of a vector $\bx$ (so $\wt(\bx)=|\supp(\bx)|$)
and $\supp({\cal D})=\cup_{\bx\in{\cal D}}\supp(\bx)$ 
for a subset ${\cal D}\in{\mathbb F}_q^n$. 
Let $(B_i,0\le i\le n)$ be the weight distribution of $D$.
The following fact, proved by MacWilliams \cite{mac63}, underlies
the combinatorial duality of coding theory:
\begin{equation}\label{eq:bin}
\sum_{i=0}^w B_i{n-i\choose n-w}=\sum_{{\cal D}\subseteq D\atop 
|\supp({\cal D})|\le w} |{\cal D}| \quad(0\le w\le n).
\end{equation}
In particular, this immediately implies the MacWilliams identities 
\cite{mac63}. Denoting ${\cal B}_w=\sum_{i=0}^w B_i{n-i\choose n-w},$
we obtain
\[
\sum_{i=0}^nB_ix^{n-i}y^i=\sum_{w=0}^n{\cal B}_w(x-y)^{n-w}y^w.
\]
Binomial moments of the weight distribution of codes were studied
extensively in \cite{ash99} and some other related works (see \cite{ash99}
for a discussion and complete bibliography). One of the reasons for
this interest is that while any particular weight component $B_i$
can be small relative to the code size, the numbers ${\cal B}_i$ can not.
The probability of undetected error (\ref{eq:Pud-classical}) can be written 
in the form
\[
P_{ue}(C,p)=\sum_{w=1}^n({\cal B}_w-1)\Big(1-{qp\over q-1}\Big)^{n-w}
\Big({p\over q-1}\Big)^w;
\]
thus lower bounds on ${\cal B}_w$ are helpful for bounding $P_{ue}(C,p)$
below.

For a quantum code $Q$ one can generalize definition (\ref{eq:bin})
by looking at error operators $E$ whose supports are of restricted
size, the support $\supp(E)$ being the subset $\{e\in\{1,2,\dots,n\}|
\tau_e\ne I_2\}.$ Then we arrive at the following weight enumerators
for $Q$:
\begin{eqnarray*}
{\cal B}_w&=&{1\over K^2}\sum_{E\in {\cal E}_n\atop |\supp(E)|\le w}
\tr^2(EP)\\
{\cal B}_w^\bot&=&{1\over K}\sum_{E\in {\cal E}_n\atop |\supp(E)|\le w}
\tr(EPEP).
\end{eqnarray*}
The generating functions of these numbers, ${\cal B}(x,y)$ and
${\cal B}^\bot (x,y)$ were studied in \cite{ref 2} and called unitary
weight enumerators. As in (\ref{eq:bin}), it is immediate that
\[
{\cal B}_w=\sum_{i=0}^w B_i{n-i\choose n-w}\quad
{\cal B}_w^\bot=\sum_{i=0}^w B_i^\bot{n-i\choose n-w},
\]
which is a result in \cite{ref 2}. The MacWilliams equation 
(\ref{mc}) also follows immediately by the original proof in \cite{mac63}.
Also, 
\[
B(x,y)={\cal B}(x-y,y)\quad B^\bot(x,y)={\cal B}^\bot(x-y,y);
\]
thus the probability of undetected error equals
\[
P_{ue}(Q,p)=\sum_{i=0}^n({\cal B}_i-{\cal B}_i^\bot)\big({p\over 3}\Big)^i
\Big(1-{4p\over 3}\Big)^{n-i}.
\]
So to bound $P_{ue}(Q,p)$ below we could first derive lower
estimates on $({\cal B}_i-{\cal B}_i^\bot)$ following the ideas of
\cite{ash99}. However, in part 2 we choose to work with the
functions (\ref{eq:stabilizer2}),\,(\ref{eq:pud-non1}) as a whole.

\appendix

\section{Appendix}
We precede the proofs of Lemmas \ref{lemma:int} and \ref{vv_vv}
with two other useful facts.
In the proofs below we repeatedly interchange the order of
integration. Obviously, all the necessary assumptions on 
the measures ($\sigma$-additivity, completeness) 
for the Fubini theorem to hold true are in place.

\begin{lemma}
\label{lemma:proj}
Let $G$ be a compact group and $\pi$ be a unitary representation 
with operators acting on a linear space $W$. Let $\mu$ be the
Haar measure on $G$. Then
$$
P:=\int_{g\in G} \pi_g d\mu(g)
$$
is an orthogonal  projection on $W$.
\end{lemma}
%{\sc Proof of Lemma \ref{lemma:proj}}.
\Proof
It suffices to show that $P^2=P$ and $P^\ast =P$. 
\begin{align*}
P^2& = \int_{h\in G} \pi_h\int_{g\in G} \pi_g d \mu(g) d \mu(h)\\
& = \int_{h\in G} \int_{g\in G} \pi_{hg}  d \mu(g) d \mu(h)=\int_{h\in G} 
\int_{g\in G} \pi_g  d \mu(g) d \mu(h) \\
& = P 
\end{align*}
Since $\pi$ is a unitary representation, $\pi_g^\ast =\pi_{g^{-1}}$. Hence
$$
P^\ast =\int_{g\in G} \pi^\ast _g d \mu(g) = \int_{g\in G} 
\pi_{g^{-1}} d \mu(g) = 
\int_{g\in G}\pi_g d \mu(g)=P.
$$
\QED

As in the proof of Theorem \ref{thm:pud-non1}, in the following
lemmas we perform
calculations in ${\mathbb C}^{K}$ instead of $Q$.
Define the inner product of matrices $A$ and $B$ as follows
\begin{equation}
\label{eq:inner-product}
\langle A,B\rangle =\tr(A^\ast B).
\end{equation}

\begin{lemma}\label{lemma:appendix2}
Let $L$ be the space of linear operators on ${\mathbb C}^K$ and $U$
be a unitary matrix. Then the equality 
\[
A\ten B=U^\ast A U\ten U^\ast B U \qquad(A, B\in L)
\]
holds true for any unitary matrix $U$ 
if and only if $A\ten B$ is contained in the subspace $W\subset L\ten L$ 
generated by
\[
\widehat{I}:=I\ten I=\sum_{i,j=1}^K \bv_i\bv_i^\ast 
\ten \bv_j\bv_j^\ast  \mbox{ and } J=
\sum_{i,j=1}^K \bv_i\bv_j^\ast \ten \bv_j\bv_i^\ast .
\]
\end{lemma}
\Proof The proof is at times sketchy, however we only omit
routine calculations. 
Let $\pi$ be the representation of $U(K)$ acting on $L\ten L$ as follows
$$
\pi_U(A\ten B)=U^\ast AU\ten U^\ast BU.
$$
We need to prove that 
$\pi_U(A\ten B)=A\ten B$ for all $U\in U(K)$ if and only if
$A\ten B\in W.$

Let us begin with the ``if'' part. It suffices to prove that 
$\pi_U$ acts identically on  $\widehat{I}$ and $J.$ 
The first of these is obvious. 
For the second, let us introduce the canonical
isomorphism
\begin{align*}
\phi:\qquad L\ten L&\to L\ten L\\
A\ten B&\mapsto B\ten A.
\end{align*}
Let us compute $\pi_U(J)$ as follows:
\begin{align*}
\pi_U(J)&=\sum_{i,j}U^\ast \bv_i\bv_j^\ast U\ten U^\ast \bv_j\bv_i^\ast U\\
&= \sum_{i,j} U^\ast \bv_i\ten 
\bv_j^\ast U\ten U^\ast \bv_j\ten \bv_i^\ast  U\\
&\stackrel{\phi}{\longrightarrow } \sum_{i,j}U^\ast \bv_i\ten 
\bv_i^\ast U\ten U^\ast \bv_j\ten \bv_j^\ast U\\
& = \sum_{i} U^\ast \bv_i\bv_i^\ast U\ten \sum_j U^\ast \bv_j\bv_j^\ast U\\
&= \sum_{i,j}\bv_i\bv_i^\ast \ten \bv_j\bv_j^\ast\\ 
&\stackrel{\phi^{-1}}{\longrightarrow} 
\sum_{i,j}\bv_i\bv_j^\ast \ten \bv_j\bv_i^\ast .
\end{align*}

Let us prove the ``only if'' part. 
Consider the group $G=DS_{K}$ where $D$ is the group of all diagonal matrices 
with diagonal elements from the set $\{ \pm1, \pm i\}$  and
$S_{K}$ is the symmetric group. Clearly, $G<U(K)$.
Let
$$
\chi={1\over |G|} \sum_{g\in G} \pi_g.
$$ 
By Lemma \ref{lemma:proj}, $\chi$ is an orthogonal projection on $G$. 
We will prove that the dimension of its image is 3. After that we will
present an operator $\bw\in \im \chi$ that is invariant under the
action of $G$ but is not fixed by $U(K)$. This will imply that the 
dimension of the subspace of $L\ten L$ fixed by 
$U(K)$ equals 2; hence by the above this subspace is $W.$

Let us find $\dim\im(\chi)=\tr(\chi)=\bu_i^\ast\chi\bu_i,$
where $(\bu_i,\,i=1,\dots,K^4)$ is any orthonormal basis of $L\ten L.$ 
For instance, take
$$
\bu_{ijkl}=\bv_i\bv_j^\ast \ten \bv_k\bv_l^\ast.
$$
Then
$$
\tr\Big({1\over |G|}\sum_{g\in G}\pi_g\Big) ={1\over |G|}\sum_{i,j,k,l}
\Big\langle \bv_i\bv_j^\ast \ten \bv_k\bv_l^\ast , 
\sum_{g\in G} \pi_g
(\bv_i\bv_j^\ast \ten \bv_k\bv_l^\ast )\Big\rangle,
$$
where $\langle\cdot,\cdot\rangle$ is the Hermitian inner product
on $L\ten L$,
\begin{equation}
\label{ten_scal_prod}
\langle A\ten B,C\ten D\rangle =\tr(AC^\ast\ten BD^\ast).
\end{equation} 

Consider an element $g\in G,\, g=\mbox{diag}(\bs)\sigma,$
where  $\bs\in \{\pm 1, \pm i \}^K$ and $\sigma\in S_K$.
It is easy to see that 
$$
\pi_g(\bv_i\bv_j^\ast \ten \bv_k\bv_l^\ast )=s_i s_j^\ast s_ks_l^\ast 
\bv_{\sigma(i)} 
\bv_{\sigma(j)}^\ast  \otimes \bv_{\sigma(k)} \bv_{\sigma(l)}^\ast .
$$
Since the basis is orthogonal, we have
$$
\tr \Big({1\over |G|}\sum_{g\in G}\pi_g\Big) ={1\over |G|}
\sum_{i,j,k,l}\Big\langle \bv_i\bv_j^\ast \ten \bv_k\bv_l^\ast , \sum_{g\in
\widehat{G}(i,j,k,l)} 
\pi_g
(\bv_i\bv_j^\ast \ten \bv_k\bv_l^\ast )\Big\rangle ,
$$
where $\widehat{G}(i,j,k,l)<G$ is a subgroup formed by the elements 
$g=\mbox{diag}(\bs)\,\sigma,$ where $\sigma$ has
fixed points $i,j,k,l$ as fixed points. 
The inner product under the sum
is nonzero only in the following three cases:\\[1mm]
a) $i=j, k=l, i\not =k$\\
b) $i=l, j=k, i\not =k$\\
c) $i=j=k=l$. \\[1mm]
It is easy to check that in each of these cases the sum 
equals $|G|$. Thus we have 
$$
\tr \Big({1\over |G|}\sum_{g\in G}\pi_g\Big) =3.
$$
To complete the proof notice that the element 
$$
\bw=\sum_{i}\bv_i\bv_i^\ast \ten \bv_i\bv_i^\ast \in W
$$
is invariant under the action of $G$ but not of $U(K)$. 
Indeed
$$
\pi_g(V)=\sum_{i=1}^K \bv_{\sigma(i)}\bv_{\sigma(i)}^*\otimes 
\bv_{\sigma(i)}\bv_{\sigma(i)}^*=V.
$$
 It is easy to check 
that $V$ is not fixed under the action of the unitary $K\times K$ matrix
\begin{equation*}
U={1 \over \sqrt{2}} 
\begin{bmatrix}
1&1& 0&\dots &0\\
1 & -1 &0 &\dots & 0\\
0&0 & \sqrt{2}&\dots  &0 \\
\vdots& \vdots&\vdots & \ddots&\vdots\\
0&0 &0 &\dots & \sqrt 2 
\end{bmatrix}
\end{equation*}
For instance for $K=2$ we have 
$$U={1 \over \sqrt{2}} \left[ \begin{array}{cc}
1 & 1\\
1 & -1
\end{array} 
\right],
$$
and it is straightforward to see that 
\begin{align*} 
U\bv_1\bv_1^\ast U\ten U\bv_1\bv_1^\ast U &+ 
U\bv_2\bv_2^\ast U\ten U\bv_2\bv_2^\ast  U\\
&= (\bv_1+\bv_2)(\bv_1+\bv_2)^{\ast}\otimes 
(\bv_1+\bv_2)(\bv_1+\bv_2)^{\ast}\\
&\phantom{==} +  
(\bv_1-\bv_2)(\bv_1-\bv_2)^{\ast}\otimes 
(\bv_1-\bv_2)(\bv_1-\bv_2)^{\ast}\\
&\not = \bv_1\bv_1^\ast \ten \bv_1\bv_1^\ast  + 
\bv_2\bv_2^\ast \ten \bv_2\bv_2^\ast.
\end{align*}
\vskip-8mm\qed

%%%%%%%END OF LEMMA FROM APPENDIX%%%%%%%%%%%%%%%%%

\bigskip
%%%%%%LEMMA 3%%%%%%%%%%%%
\nd{\sc Proof of Lemma \ref{lemma:int}}. 

Let $\bw_i$ and $\bv_i$ be orthonormal bases of $Q$ and 
${\mathbb C}^K,$ respectively. Let $\phi:\,Q\to {\mathbb C}^K$ be the natural
isomorphism given by $\phi(\bw_i)=\bv_i.$
Denote by $\rho(U)$ the Haar measure on the unitary group $U(K)$. 
Applying $\phi,$ we can rewrite the integral in question as follows
\begin{align*}
\int_{\bv \in Q} \bw\bw^\ast  d \mu(\bv)
\stackrel{\phi}{\longrightarrow}  
&\int_{\bv \in {\mathbb C}^K} \bv\bv^\ast  d \mu(\bv) \\
& =  \int_{\bv \in {\mathbb C}^K} U^\ast  \bv\bv^\ast  U d \mu(\bv)\\
&=  \int_{U\in U(K)} d \rho(U) \int_{\bv \in {\mathbb C}^K} 
U^\ast \bv\bv^\ast U d \mu(\bv) \\
&= \int_{U\in U(K)}  U^\ast  \bv\bv^\ast  U d \rho(U),
\end{align*}
where in the last expression $\bv$ is an arbitrary fixed
basis vector. We need to compute the last integral.

Let $\pi$ be a unitary representation of $U(K)$ acting on the vector 
space of complex $K\times K$ matrices as follows
$$
\pi_U(\bv\bv^\ast )=U^\ast \bv \bv^\ast U.
$$
Consider the operator 
\begin{equation*}
\label{pi}
\int_{U\in U(K)}  U^\ast  \bv\bv^\ast  U d \rho(U)=
\int_{U\in U(K)} \pi_U(\bv\bv^\ast ) d \rho(U)
\end{equation*}
by Lemma \ref{lemma:proj} it is an orthogonal projection.
Next we show that it projects on the one-dimensional
subspace generated by $I_K$; then the last integral becomes 
easy to compute.

Note that $\pi$ is unitary with respect to the inner product
(\ref{eq:inner-product}). Indeed,
$$
\langle \pi_U(A),\pi_U(B)\rangle =\tr(U^\ast AUU^\ast BU)
=\tr(A^\ast B)=\langle A,B\rangle .
$$
   From the standard fact that 
any matrix $A'$ can be represented in the form $A'=UAW$,
where $U$ and $W$ are unitary matrices and $A$ is diagonal,
it is easy to see that the identity $U^\ast AU=A$ can hold for
any unitary matrix $U$ if and only if $A=I_K$
(this is a particular case of Lemma \ref{lemma:appendix2}). 
Therefore 
$$
\int_{U\in U(K)}  U^\ast  \bv\bv^\ast  U d \rho(U)
$$
equals the projection of $\bv^\ast \bv$ on the one-dimensional subspace 
of $W$ generated by $I_K$. 
Recalling that the projection of a vector $\bx$ on a vector $\by$ equals
$
{\langle \bx, \by\rangle \over \langle \by,\by\rangle } \by,
$
we have
$$
\int_{U\in U(K)}  U^\ast  \bv\bv^\ast  U d \rho(U)=
{\langle \bv^\ast \bv,I_K\rangle \over \langle I_K,I_K\rangle } I_K={ 
1 \over K} I_K={1 \over K} \sum_{i=1}^K \bw_i\bw_i^{\ast}.
$$
Finally, since
$$
\phi^{-1}\Big(\sum_{i=1}^K \bw_i\bw_i^{\ast}\Big) = 
\sum_{i=1}^K \bv_i\bv_i^{\ast} =P,
$$
we are done.  \qed
%%%%%%%%%%%%%%%%%%%END OF LEMMA 3%%%%%%%%%%%%%%%%%%%%%%%%%%

%%%%%%%%%%%LEMMA 4%%%%%%%%%%%%%%%%%%%%%
\bigskip\nd
{\sc Proof of Lemma \ref{vv_vv}}.

Let $L$ be the space of linear operators on ${\mathbb C}^K.$ Similarly to 
the proof of Lemma \ref{lemma:int} it can be seen that
\begin{equation}
\label{ten_int}
\int_{\bv\in {\mathbb C}^K} \bv\bv^\ast \ten \bv\bv^\ast  
d\mu(\bv)=\int_{U\in U(K)} 
U^\ast {\tilde{\bv}}{\tilde{\bv}}^\ast U\ten U^\ast {\tilde{\bv}}
{\tilde{\bv}}^\ast U d\rho(U),
\end{equation}
where $\tilde \bv$ is an arbitrary fixed basis vector.
By Lemma \ref{lemma:appendix2} the integral in (\ref{ten_int})
is a projector, with respect to the inner product
(\ref{ten_scal_prod}),
on the subspace $W\subset L\ten L$ generated by 
$\widehat I$ and $J.$
The lemma will be proved if we evaluate the projection on
this subspace of $\bu:={\tilde{\bv}}{\tilde{\bv}}^\ast \ten 
{\tilde{\bv}}{\tilde{\bv}}^\ast. $ 

To do this, we need an orthogonal basis of $W$ (note that
$\widehat I$ and $J$ are not orthogonal; indeed,
it is easy to check that 
$$
\langle \widehat{I},\widehat{I}\rangle =K^2,
 \langle J,J\rangle =K^2, \mbox{ and } \langle \widehat{I},J\rangle =K.)
$$
Therefore, let us consider the basis $\widehat I,\widehat{J}$, where
$$
\widehat{J}=J-{1\over K}\widehat{J}.
$$
Again it is easy to see that
$$
\langle \widehat{J},\widehat{J}\rangle =K^2-1 \mbox{ and } 
\langle \widehat{I},\widehat{J}\rangle =0.
$$
Thus, $\widehat{I}$ and  $\widehat{J}$ form an orthogonal 
basis of $W$. To compute the projection of $\bu$ on $W$,
let us first project it on the basis directions: 
\begin{eqnarray*}
\langle \widehat{I}, \bu\rangle &=&
\tr\biggl(\sum_{i,j}(\bv_i\bv_j^\ast \ten 
\bv_i\bv_j^\ast )({\tilde{\bv}}{\tilde{\bv}}^\ast \ten 
{\tilde{\bv}}{\tilde{\bv}}^\ast )\biggr)
\nonumber \\
&=& \sum_{i,j} \tr(\bv_i\bv_j^\ast {\tilde{\bv}}{\tilde{\bv}}^\ast )
\tr(\bv_i\bv_j^\ast {\tilde{\bv}}{\tilde{\bv}}^\ast )\nonumber\\
&=&1; \nonumber\\
\langle \widehat{J},\bu\rangle &=&
\Big(1-{1\over K}\Big).
\end{eqnarray*}
Thus, the projection of $\bu $ on $W$ equals
$$
\int_{\bv\in {\mathbb C}^K} \bv\bv^\ast \ten \bv\bv^\ast  d\mu(\bv)
={\langle 
\widehat{I},\bu\rangle  \over \langle \widehat{I},\widehat{I}\rangle } 
\widehat{I} 
+{\langle \widehat{J},\bu\rangle \over \langle \widehat{J},\widehat{J}
\rangle }\widehat{J}=
{1\over K(K+1)}(\widehat{I}+J).
$$
This completes the proof.\QED
%%%%%%%%%END OF LEMMA 4%%%%%%%%%%%%%%%%%%%%%%%%%


{\footnotesize
\begin{thebibliography}{99}
\bibitem{ash99}A.~Ashikhmin and A.~Barg, ``Binomial moments of the 
distance distribution: {B}ounds and applications,'' 
IEEE Trans. Inform. Theory, vol. {45}, number 2, pp. 438--452, 1999.

\bibitem{ABKL} A. Ashikhmin, A. Barg, E. Knill, and S. Litsyn,
``Quantum error detection, II: Bounds'', {\em IEEE Trans. Inform.
Theory}, submitted.

\bibitem{CCKS}
A. R. Calderbank, P. J. Cameron, W. M. Kantor, and
J. J. Seidel, ``${\mathbb Z}_4$-Kerdock codes, orthogonal
spreads, and extremal euclidean line-sets,''
{\em Proc. London Math. Soc.}, vol. 75, number 3, pp. 436--480, 1997.
  
\bibitem{Cal geom} A.R. Calderbank, E.M. Rains, P.W. Shor and N.J.A. Sloane,  
 ``Quantum error correction and orthogonal geometry,'' 
 {\em Phys. Rev. Lett.}, vol. 78, pp. 405-409, 1997.  

\bibitem{Cal gf} 
 A.R. Calderbank, E.M. Rains, P.W. Shor and N.J.A. Sloane, ``Quantum errors 
correction via codes 
over $GF(4)$,''{\it IEEE Trans. Info. Theory}, vol. 44, pp.1369 --1387, 1998. 


\bibitem{del73}
P.\,Delsarte,  {\em An algebraic approach to the association schemes of
coding theory}, Philips Research Reports Supplements, No. 10, 1973.


%\bibitem{ekert96} A. Ekert and C. Macchiavello, "Error correction in quantum
%   communication," {\em  Phys. Rev. Lett.}, vol. 77, pp. 2585-2588,
%   1996.

\bibitem{Gottesman} D. Gottesman, ``A class of quantum 
error-correcting codes saturating the quantum Hamming bound,'' 
{\em  Phys. Rev. A}, vol.54, pp. 1862-1868, 1996. 

\bibitem{holevo} A. S. Holevo, {\em Probabilistic and 
Statistical Aspects of Quantum Theory}, Amsterdam: North Holland
Publ. Co., 1982. 

\bibitem{KK} T.Kl{\o}ve and V. Korzhik, {\em Error Detecting Codes},
Dordrecht: Kluwer, 1995.

\bibitem{kos} A. I. Kostrikin and Yu. I. Manin,
{\em Linear Algebra and Geometry}, Gordon and Breach Science Pub., 1989.

\bibitem{knill}
E. Knill and R. Laflamme, ``A theory of quantum error correcting codes,''
{\em Phys. Rev. A}, vol. 55, pp. 900-911, 1997.

\bibitem{mac63}
F.~J. MacWilliams, ``A theorem in the distribution of weights in a
  systematic code'', {\em Bell Syst. Techn. Journ.}, vol. {42} 
pp. 79--94, 1963.
 
\bibitem{peres}
A.\,Peres, {\em Quantum Theory: Concepts and Methods}, 
Dordrecht: Kluwer, 1995.

\bibitem{prasolov}
V. V. Prasolov, {\em Problems and Theorems in Linear Algebra},
Providence, RI: Amer Math. Soc., 1994.

\bibitem{preskill} J. Preskill, {\em Quantum Information and
Computation}, Lecture Notes for Physics 229, available from
http://www.theory.caltech.edu/~preskill/ph229.

\bibitem{ref 2}
E.M.~Rains, ``Quantum weight enumerators,'' 
{\it IEEE Trans. Info. Theory}, vol. 
44, pp.1388--1394, 1998.

\bibitem{ref 12}
E.M.~Rains, ``Quantum shadow enumerators,'' LANL e-print
quant-ph/961101.

\bibitem{ref 1}
P.W.~Shor and R. Laflamme, ``Quantum analog of the MacWilliams identities
in classical coding theory,'' {\em Phys. Rev. Lett.}, vol. 78, pp. 1600-1602, 
1997.

%\bibitem{shor2} P.W.~Shor, ``Polynomial-time algorithms for prime
%factorization and discrete logarithms on a quantum computer,''
%{\em Proceedings of the 35th Annual Symposium on the 
%Foundations of Computer 
%Science}, S.Goldwasser, Editor, IEEE Computer Society Press, 
%Los Alamitos, CA, p.124, 1994.

\bibitem{shor1} P.W.~Shor, ``Scheme for reducing decoherence in quantum
memory,'' Phys. Rev. A,
{\bf 52}, p. 2493, 1995.

\bibitem{steane1} 
A. M. Steane,  "Simple quantum error correcting codes," 
{\em Phys. Rev. Lett.}, 
vol. 77, pp. 793-797, 1996.

%\bibitem{steane2} 
%A. M. Steane, "Multiple particle interference and quantum
%   error correction," {\em Proc. Roy. Soc. London A}, vol. 452, pp.
%   2551-2577, 1996.
 
%\bibitem{wooters}
%W.~K.~Wooters and W.~H.~Zurek, ``A single quantum cannot be cloned,''
%{\it Nature}, {\bf 299}, p.802, 1982.

\end{thebibliography}
}
\end{document}